\documentclass[twocolumn,showpacs,showkeys]{revtex4}
\usepackage{graphicx}
\input{psfig.sty}

\newcommand{\bastar}{\begin{eqnarray*}}
\newcommand{\eastar}{\end{eqnarray*}}
\newskip\humongous \humongous=0pt plus 1000pt minus 1000pt

\newif\ifdtup

\relax
\newcommand{\be}{\begin{equation}}
\newcommand{\ee}{\end{equation}}
\newcommand{\bea}{\begin{eqnarray}}
\newcommand{\eea}{\end{eqnarray}}

\newcommand{\hn}{\hat n}

\newcommand{\dfrac}{\displaystyle\frac}
\newcommand{\nn}{\nonumber}

\begin{document}
\title  {D-type Vortex and N-type Vortex in Two-gap Superconductor}
\bigskip
\author{Y. M. Cho}
\email{ymcho@yongmin.snu.ac.kr}
\author{Pengming Zhang}
\email{zhpm@phya.snu.ac.kr}
\affiliation{Center for Theoretical Physics and School of Physics \\
College of Natural Sciences, Seoul National University, Seoul
151-742, Korea  \\}

\begin{abstract}

We show that the two-gap superconductor has two types of
non-Abrikosov magnetic vortex, the D-type
which has no concentration of the condensate
at the core and the N-type which has a non-vanishing
concentration of the condensate at the core, which may
carry $4\pi$-flux, $2\pi$-flux, or a fractional flux,
depending on the parameters of the potential.
Furthermore, we show that the non-Abrikosov vortex
is described by two types of topology. The integer flux vortex has
the non-Abelian topology $\pi_2(S^2)$, but the fractional flux
vortex has the Abelian topology $\pi_1(S^1)$.
We show that the inclusion of the Josephson interaction
does not affect the existence of the magnetic vortex,
but alter the shape of the vortex drastically.
\end{abstract}
\pacs{74.20.-z, 74.20.De, 74.60.Ge, 74.60.Jg, 74.90.+n}
\keywords{non-Abrikosov vortex, fractional magnetic flux
in two-gap superconductor}

\maketitle

The Abrikosov vortex in one-gap superconductors and similar
ones in Bose-Einstein condensates and superfluids have played
a very important role in condensed matter physics \cite{abri}.
But the recent advent of two-component Bose-Einstein condensates
and two-gap supercondectors \cite{bec,sc}
has opened up an exciting new possibility
for us to consrtuct far more interesting topological
objects in laboratories \cite{ijpap,prb05,ruo,pra05,baba,cm2}.
It has bee shown that the two-gap supercondector can
admit a non-Abrikosov vortex which has the non-Abelian
$\pi_2(S^2)$ topology \cite{ijpap,pra05,cm2}.
{\it The purpose of this Letter is to show that the two-gap supercondector
actually allows two types of non-Abrikosov vortex,
D-type and N-type, whose magnetic flux can be anywhere between
$4\pi/g$ and zero.} And this holds true even when the interband
Josephson interaction is present. The reason for this is that
the magnetic vortex in two-gap superconductor
has two types of boundary condition at the core.

In mean field approximation
the free energy of the two-gap
superconductor could be expressed by \cite{ijpap,baba}
\bea
&{\cal H} = \dfrac{\hbar^2}{2m_1} |(\mbox{\boldmath $\nabla$}
+ ig \mbox{\boldmath $A$}) \tilde \phi_1|^2
+\dfrac{\hbar^2}{2m_2} |(\mbox{\boldmath $\nabla$}
+ ig \mbox{\boldmath $A$}) \tilde \phi_2|^2 \nn\\
&+ \tilde V(\tilde \phi_1,\tilde \phi_2)
+ \dfrac{1}{2} (\mbox{\boldmath $\nabla$}
\times \mbox{\boldmath $A$})^2,
\label{scfe1}
\eea
where $\tilde V$ is the effective potential.
We choose the potential to be the most general quartic
potential which has the $U(1)\times U(1)$ symmetry,
\bea
&\tilde V =\dfrac{\tilde{\lambda}_{11}}2|\tilde{\phi}_1|^4
+\tilde{\lambda}_{12}|\tilde{\phi}_1|^2|\tilde{\phi}_2|^2
+\dfrac{\tilde{\lambda}_{22}}2| \tilde{\phi}_2|^4 \nn\\
&-\tilde{\mu}_1|\tilde{\phi}_1|^2
-\tilde{\mu}_2|\tilde{\phi}_2|^2,
\label{scpot1}
\eea
where $\tilde{\lambda}_{ij}$ are the quartic coupling
constants and $\tilde \mu_i$ are the chemical potentials of
$\tilde \phi_i$ ($i=1,2$). The Josephson interaction which breaks the
$U(1)\times U(1)$ symmetry down to $U(1)$ will be discussed separately
in the following.

With the normalization
of $\tilde \phi_1$ and $\tilde \phi_2$ to $\phi_1$ and $\phi_2$,
\bea
\phi_1=\dfrac \hbar {\sqrt{2m_1}}\tilde{\phi}_1,\;\;\;\;
\phi_2=\frac \hbar {\sqrt{2m_2}}\tilde{\phi}_2.
\eea
one can simplify the above Hamiltonian,
\bea
&{\cal H} = |(\mbox{\boldmath $\nabla$}
+ ig \mbox{\boldmath $A$}) \phi|^2 + V(\phi_1,\phi_2)
+ \dfrac{1}{2} (\mbox{\boldmath $\nabla$}
\times \mbox{\boldmath $A$})^2,
\label{scfe2}
\eea
where $V$ is the normalized potential,
\bea
&V =\dfrac{\lambda_{11}}{2}|\phi_1|^4+\lambda_{12}|\phi_1|^2
|\phi_2|^2+\dfrac{\lambda_{22}}{2}|\phi_2|^4 \nn\\
&-\mu_1|\phi_1|^2-\mu_2|\phi_2|^2.
\label{scpot2}
\eea
The potential has the following types of vacuum: \\
A. Type I: Integer flux vacuum
\bea
\left( \begin{array}{l} <|\phi_1|> \\
<|\phi_2|> \end{array} \right)
=\left( \begin{array}{l} \sqrt{\mu_1/\lambda_{11}} \\
0
\end{array} \right).
\label{vacb}
\eea
This is possible when we have one of the following
three cases,
\bea
&&(a)~~~~0< \lambda_{12},~~~\dfrac{\lambda_{12}}{\lambda_{22}}
<\dfrac{\lambda_{11}}{\lambda_{12}} \leq \dfrac{\mu_1}{\mu_2}, \nn\\
&&(b)~~~~0< \lambda_{12},~~~\dfrac{\lambda_{11}}{\lambda_{12}}
<\sqrt{\dfrac{\lambda_{11}}{\lambda_{22}}}
<\dfrac{\mu_1}{\mu_2}, \nn\\
&&(c)~~~~0< \lambda_{12},~~~\dfrac{\lambda_{11}}{\lambda_{12}}
=\dfrac{\lambda_{12}}{\lambda_{22}}<\dfrac{\mu_1}{\mu_2}.
\eea
We call this integer flux vacuum because, as we will see,
the magnetic vortex with this type of vacuum has
an integer flux. \\
B. Type II: Fractional flux vacuum
\bea
&\left( \begin{array}{l}
<|\phi_1|> \\ <|\phi_2|> \end{array} \right)
=\left( \begin{array}{l} \hat{\phi}_1 \\ \hat{\phi}_2
\end{array} \right), \nn\\
&\hat \phi_1^2 = \dfrac{\mu_1\lambda_{22}-\mu_2\lambda_{12}}
{\lambda_{11} \lambda_{22} - \lambda_{12}^2},
~~~\hat \phi_2^2 = \dfrac{\mu_2\lambda_{11}
-\mu_1\lambda_{12}}{\lambda_{11} \lambda_{22} - \lambda_{12}^2}.
\label{vaca}
\eea
This is possible when we have one of the following
three cases,
\bea
&&(a)~~~~\lambda_{12} <0,~~~\dfrac{|\lambda_{12}|}{\lambda_{22}}
<\dfrac{\lambda_{11}}{|\lambda_{12}|}, \nn\\
&&(b)~~~~0< \lambda_{12},~~~\dfrac{\lambda_{12}}{\lambda_{22}}
<\dfrac{\mu_1}{\mu_2}<\dfrac{\lambda_{11}}{\lambda_{12}}, \nn\\
&&(c)~~~~\lambda_{12} =0.
\eea
We call this fractional flux vacuum because
the magnetic vortex with this type of vacuum has
a fractional flux. \\
C. Type III: Degenerate vacuum
\bea
\mu_1 <|\phi_1|>^2 + \mu_2 <|\phi_2|>^2 = \dfrac{\mu_1 \mu_2}{\lambda_{12}}.
\label{vacc}
\eea
This is what we have when
\bea
\dfrac{\mu_1}{\mu_2}=\dfrac{\lambda_{12}}{\lambda_{11}}
=\dfrac{\lambda_{22}}{\lambda_{12}}.
\eea
This includes the special (and familiar) case
\bea
&\lambda_{11}=\lambda_{12}=\lambda_{22}=\lambda,
~~~\mu_1=\mu_2=\mu.
\eea
In this case the potential (\ref{scpot2}) has the full $SU(2)$ symmetry.
Notice that the potential (\ref{scpot2}) has no vacuum
when
\bea
\lambda_{12}< 0,~~~\dfrac{\lambda_{11}}{|\lambda_{12}|}
\leq \dfrac{|\lambda_{12}|}{\lambda_{22}}.
\eea
All other cases can be reduced to one of the above cases
by re-labelling $\phi_1$ and $\phi_2$.

With
\bea
\phi =\dfrac{1}{\sqrt 2} \rho \xi~~~~~({\xi}^{\dagger}\xi = 1),
~~~~~\hat n = \xi^{\dagger} \vec \sigma \xi,
\label{ndef}
\eea
we can obtain the following equation of motion from
the Hamiltonian (\ref{scfe2}) \cite{ijpap,prb05,cm2}
\begin{eqnarray}
&\partial^2\rho -\Big( \dfrac 14(\partial _\mu \hn)^2+g^2(A_\mu
-\dfrac 12C_\mu )^2\Big) \rho   \nonumber \\
&= \Big[\dfrac \lambda2 (\rho ^2- \bar \rho^2)
+(\dfrac{\alpha}2 \rho^2-\gamma) n_3
+\dfrac{\beta}2 \rho^2 n_3^2 \Big] \rho, \nonumber \\
&\hat{n}\times \partial ^2\hat{n}+2\dfrac{\partial _\mu \rho }%
\rho \hat{n}\times \partial _\mu \hat{n}-\dfrac 2{g\rho
^2}\partial _\mu F_{\mu \nu }\partial _\nu \hat{n}  \nonumber \\
&=\Big(2\gamma -(\dfrac{\alpha}{2}+\beta n_3) \rho^2 \Big)
\hat{k}\times \hat{n},  \nonumber \\
&\partial _\mu F_{\mu \nu }=j_\nu =g^2\rho ^2\Big(A_\nu
-\dfrac12 C_\nu \Big),
\label{sceq}
\end{eqnarray}
where
\begin{eqnarray}
&C_\mu =\dfrac{2i}g\xi ^{\dagger }\partial _\mu \xi,
~~~~~\bar \rho^2=\dfrac{2\mu}{\lambda}, \nn\\
&\lambda =\dfrac{\lambda_{11}+\lambda _{22}+2\lambda _{12}}4, \nn\\
&\alpha =\dfrac{\lambda_{11}-\lambda_{22}}2,
~~~\beta =\dfrac{\lambda_{11}+\lambda_{22}-2\lambda_{12}}4, \nn\\
&\mu =\dfrac{\mu_1+\mu_2}2,~~~\gamma =\dfrac{\mu _1-\mu _2}2, \nn
\end{eqnarray}
and $\hat k=(0,0,1)$. This is the equation for two-gap superconductor
which allows a large class of interesting topological
objects, straight magnetic vortex, helical magnetic vortex,
and magnetic knot, all with $4\pi/g$-flux, $2\pi/g$-flux,
or a fractional flux.

To discuss the vortex solution let
$(\varrho,\varphi,z)$ be the cylindrical coordinates and
choose the ansatz
\begin{eqnarray}
&\rho =\rho (\varrho ),~~~~~\xi =\left(
\begin{array}{c}
\cos \dfrac{f(\varrho )}2\exp (-in\varphi ) \\
\sin \dfrac{f(\varrho )}2
\end{array} \right) , \nn\\
&A_\mu =\dfrac ngA(\varrho)\partial_\mu \varphi.
\label{svans}
\end{eqnarray}
With this one can obtain the vortex solution by solving
(\ref{sceq}). To do this, we have to fix the boundary
conditions. At the core the smoothness allows (for $n=1$)
two types of boundary condition \cite{ijpap,prb05,cm2}: \\
A. Dirichlet boundary condition
\bea
\rho(0)=0,~~~\dot{\rho}(0) \neq 0,~~~A(0)=-1.
\label{dbc}
\eea
B. Neumann boundary condition
\bea
\rho(0)\neq 0,~~~\dot{\rho}(0)=0,~~~A(0)=0.
\label{nbc}
\eea
This is a new feature of two-gap superconductor,
which we do not have in ordinary superconductor.

At the infinity all fields must assume the
vacuum values. In particular, the electromagnetic current
must vanish.
This means that we must have
\begin{eqnarray}
&\rho(\infty)= \sqrt{2(<|\phi_1|>^2+<|\phi_2|>^2)},\nn\\
&\cos f(\infty)=\dfrac{<|\phi_1|>^2-<|\phi_2|>^2}
{<|\phi_1|>^2+<|\phi_2|>^2}, \nn\\
&A(\infty)=\dfrac {<|\phi_1|>^2} {<|\phi_1|>^2+<|\phi_2|>^2}.
\label{bcinf}
\end{eqnarray}
Notice that for the integer flux vacuum we have $A(\infty)=1$,
but for the fractional flux vacuum $A(\infty)$ becomes
fractional.

The existence of two types of boundary conditions
in two-gap superconductor has an important impact.
To understand this notice that the magnetic flux of vortex
is given by
\begin{eqnarray}
\Phi = \oint A_\mu dx^\mu = \dfrac {2 \pi n}{g}
\big( A(\infty)-A(0) \big).
\label{mflux}
\end{eqnarray}
So, the magnetic flux becomes fractional
when $A(\infty)$ is fractional.
As importantly, when $A(0)=-1$ the flux becomes $4\pi/g$
with $A(\infty)=1$. This was impossible in ordinary superconductor.

\begin{figure}[t]
\includegraphics[scale=0.52]{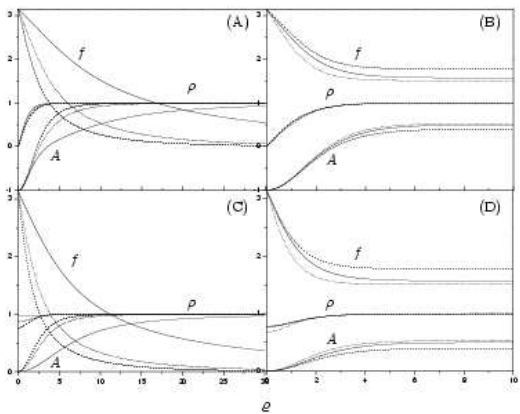}
\caption{The non-Abrikosov vortices in two-gap supercomductor.
In (A) and (B) the D-type vortices with $4\pi/g$ flux
and fractional flux are shown with different $\alpha,~\beta,~\gamma$.
In (C) and (D) the N-type vortex with $2\pi/g$ flux
and fractional flux are shown with different $\alpha,~\beta,~\gamma$.
Here we have put $n=1$ and $\lambda/g^2=2$,
and the unit of the scale is $1/{\bar \rho}$.}
\label{scsv}
\end{figure}

With the ansatz (\ref{svans}) we obtain the following vortex solutions. \\
{\bf A. $4\pi/g$-flux vortex}: With the Dirichlet boundary condition
at the core and the integer flux vacuum at the infinity \cite{ijpap},
\begin{eqnarray}
&\rho (0)=0,~~~~~\rho (\infty )
=\sqrt{\dfrac{2(\mu +\gamma )}{(\lambda +\alpha+\beta )}}, \nn\\
&f(0)=\pi,~~~~~f(\infty ) =0, \nn\\
&A(0)=-1,~~~~~A(\infty )=1,
\end{eqnarray}
we find the solutions with the
different parameters which are shown in Fig. 1A.
We call this a D-type vortex.
Both $\phi_1$ and $\phi_2$ start from zero at the core.
However, notice that $\phi_1$ approches the finite vacuum value
but $\phi_2$ approches zero at the infinity. So $\phi_2$
has a maximum concentration at a finite distance
from the core. This is a generic feature of a D-type vortex.
The magnetic flux of the vortex is given by
\begin{eqnarray}
\Phi =\int F_{\varrho \varphi}d^2x
=\int \partial_\varrho A_\varphi d^2x=\dfrac{4\pi n}g.
\end{eqnarray}
This has a non-Abelian topology. To see this notice that
$\hn$ defines a non-trivial mapping $\pi_2(S^2)=n$ from the compactifed
$xy$-plane $S^2$ to the $CP^1$ space $S^2$.
Clearly this is non-Abelian. \\
{\bf B. $2\pi/g$-flux vortex}: With the Neumann boundary condition
at the core and the integer flux vacuum at the infinity \cite{cm2},
\begin{eqnarray}
&\dot \rho(0)=0,~~~~~\rho (\infty)
=\sqrt{\dfrac{2(\mu +\gamma )}{(\lambda +\alpha +\beta )}}, \nn\\
&f(0)=\pi,~~~~~f(\infty ) =0,  \nn\\
&A(0)=0,~~~~~A(\infty )=1,
\end{eqnarray}
we find the solutions with the
different parameters shown in Fig. 1C.
We call this a N-type vortex.
In this case $\phi_1$ behavior is the same as before.
But notice that $\phi_2$ has a maximum concentration at the core,
and approaches zero at the infinity.
This is a generic feature of a N-type vortex.
The magnetic flux of the vortex is given by
\begin{eqnarray}
\Phi =\int F_{\varrho \varphi}d^2x
=\int \partial_\varrho A_\varphi d^2x=\dfrac{2\pi n}g,
\end{eqnarray}
which is same as that of Abrikosov vortex.
But the topology of the $CP^1$ field $\hn$ is the same as
the $4\pi/g$-flux vortex, $\pi_2(S^2)=n$. The reason why there exist
two vortices which have different magnetic flux but have the same topology
is because the magnetic flux is determined by
the boundary condition $A(\infty)-A(0)$, not by the topology.
The topology assures the quantization of the flux, but does not
determine the magnitude of the flux\\
{\bf C. Fractional flux vortex}: This is possible when
we have the fractional flux vacuum
at the infinity
\begin{eqnarray}
&\rho (\infty )=2\sqrt{\dfrac{2 \beta \mu -\alpha \gamma}
{4\beta \lambda-\alpha^2}},
~~~\cos f(\infty )=\dfrac{2\gamma \lambda -\alpha \mu}
{2\beta \mu-\alpha \gamma}, \nn\\
&A(\infty )=\dfrac 12 \dfrac{2(\gamma \lambda + \beta \mu)
-\alpha(\mu+\gamma)}{2\beta \mu-\alpha \gamma}.
\end{eqnarray}
At the core we can impose either the Dirichlet condition
(\ref{dbc}) or the Neumann condition (\ref{nbc}), and
obtain the D-type vortex shown in Fig. 1B or the N-type
vortex shown in Fig. 1D.
The fractional vortex is also topological,
but the topology of the fractional vortex is different from
that of integer flux vortex.
For the fractional flux vortices
the $\pi_2(S^2)$ topology of $\hn$ becomes trivial,
but in this case we still have
a $U(1)$ topology $\pi_1(S^1)$, the topology of the $U(1)$
symmetry which leaves $\hn$ invariant. And this Abelian
topology describes the topology of
the fractional flux vortex. So the topology of
the fractional flux vortices is the same as that of
the Abrikosov vortex.
An important feature of the fractional flux vortex is that
the energy per unit length of the vortex is logarithmically
divergent. This, however, does not make
the fractional flux vortex unphysical.
In laboratory setting one can observe such vortex because
one has a natural cutoff
parameter $\Lambda$ fixed by the size of the superconductor,
which can effectively make the
energy of the fractional flux vortex finite.

\begin{figure}[t]
\includegraphics[scale=0.35]{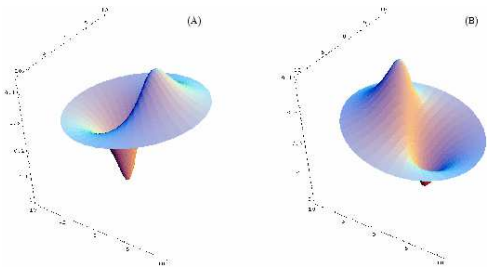}
\caption{The density profile of $|\phi_1|^2$ in (A) and
$|\phi_2|^2$ in (B) of the magnetic vortex
in the presence of Josephson interaction.
Here we have put $\bar \rho=1$, $\gamma=0.05$, $\eta = 0.25$, and
$\lambda/g^2=2$.}
\label{jphia}
\end{figure}

It must be emphasized, however, that the actual magnetic flux
of vortex depends on the two-gap superconductor
at hand because it is fixed by the parameters of
the potential which charactrizes
the superconductor. Independent of this all two-gap superconductors
have two types of vortex, D-type and N-type.
On the other hand one must keep the followings in mind.
First, the D-type vortex has more energy than the N-type vortex,
because it carries $2\pi/g$ more flux. Secondly,
the energy (per unit length) of the fractional flux vortex
is logarithmically divergent, so that they can exist only
when a cutoff parameter makes the energy finite.

It is well-known that two-gap superconductor may allow
the interband Josephson interaction \cite{leg}.
To discuss the impact of the Josephson interaction in two-gap
superconductor, we let
$\lambda_{11}=\lambda_{22}=\lambda_{12}=\lambda$
and adopt the potential which has the Josephson interaction
\bea
&V= \dfrac{\lambda}2 (|\phi_1|^2 + |\phi_2|^2)^2
- \mu_1|\phi_1|^2 - \mu_2|\phi_2|^2 \nn\\
&+\eta (\phi_1^{*}\phi_2 +\phi_1\phi_2^{*}).
\label{jpot}
\eea
Now, we choose the following ansatz for the magnetic vortex
\bea
&\rho=\rho(\varrho), \nn\\
&\xi= \Bigg( \matrix {\cos \dfrac f2 \cos \dfrac{\omega}2
\exp(-in\varphi) + \sin \dfrac f2 \sin \dfrac{\omega}2 \cr
-\cos \dfrac f2 \sin \dfrac{\omega}2 \exp(-in\varphi)
+\sin \dfrac f2 \cos \dfrac{\omega}2 } \Bigg), \nn\\
&A_\mu =\dfrac ngA(\varrho)\partial_\mu \varphi,
~~~\tan \omega= \dfrac{2\eta}{\mu_1-\mu_2}.
\label{jvans}
\eea
Notice that the ansatz is not axially symmetric.
With this we obtain two types of magnetic vortex.
The density profile of the N-type $2\pi/g$-flux vortex
is plotted in Fig.~\ref{jphia}. Notice that the vortex
can be viewed as a ``bound state" of two vortices made of
$\phi_1$ and $\phi_2$.
This confirms that the Josephson interaction does
not prevent the existence of two types of magnetic vortex.
It makes the vortex more interesting.

Finally we emphasize that all these non-Abrikosov vortices
can be twisted to form a helical vortex which is periodic in
$z$-coodinate. In particular, with the Josephson interaction,
we can construct a ``braided" helical vortex from
the above bound state of $\phi_1$ and $\phi_2$ vortices
by twisting and making it periodic in $z$-coordinate.
Perhaps more importantly, we can construct
a twisted magnetic vortex ring from the helical vortex
by smoothly bending it and connecting two periodic ends
together. And the vortex ring
becomes a stable magnetic knot whose knot topology $\pi_3(S^2)$
is fixed by the Chern-Simon index of the electromagnetic
potential \cite{ijpap,cm2}.
Because of the helical structure of the
magnetic flux the knot has two magnetic flux linked together,
one around the knot tube and one along the knot,
whose linking number is given by the knot quantum number.
And since the flux trapped inside the vortex ring
can not be squeezed out, it makes the knot stable against
collapse by providing a stabilizing repulsive force.
This makes the knot dynamically (as well as topologically) stable.

It is really remarkable that the two-gap superconductor
allows such diverse topological objects. This is because it has
a non-Abelian structure in which
the doublet $\phi$ can be treated as an $SU(2)$
doublet. The fact that the Hamiltonian (\ref{scfe2}) has
the full $SU(2)$ symmetry when $\alpha=\beta=\gamma=0$
tells that the $SU(2)$ symmetry survives
as an approximate symmetry of two-gap superconductor, which is
broken by the $\alpha$, $\beta$, and $\gamma$
interactions \cite{ijpap,prb05}.

There are other topological objects which have not been discussed
in this paper. The detailed discussion on the topological objects in
two-gap superconductor will be
presented in a separate paper \cite{sc3}.

{\bf ACKNOWLEDGEMENT}

The work is supported in part by the ABRL Program of Korea Science
and Engineering Foundation (Grant R02-2003-000-10043-0).

\end{document}